\documentclass[sigconf,nonacm]{acmart}

\usepackage{amsmath}
\usepackage{amsfonts}
\usepackage{nicefrac}

\usepackage{algorithm}
\usepackage{algorithmic}

\usepackage{array}
\usepackage{booktabs}
\usepackage{makecell}
\usepackage{tabularx}
\usepackage{multirow}

\usepackage{graphicx}
\usepackage{subcaption}
\usepackage{tikz}
\usetikzlibrary{
  arrows.meta,
  positioning,
  shapes.geometric,
  calc,
  backgrounds,
  fit
}

\usepackage{pgfplots}
\pgfplotsset{compat=1.18}

\usepackage[most]{tcolorbox}
\tcbuselibrary{breakable}
\usepackage{listings}
\usepackage{enumitem}

\usepackage{xspace}
\usepackage{xurl}
\usepackage{placeins}
\usepackage{pifont}

\usepackage{float}

\newcommand{\cmark}{\ding{51}}
\newcommand{\xmark}{\ding{55}}

\newcommand{\sys}{\textsc{ReVision}\xspace}
\newcommand{\mypara}[1]{\noindent\textbf{#1:}\ }

\AtBeginDocument{%
}

\begin{document}

\title{\sys: A Post-Hoc, Vision-Based Technique for Replacing
Unacceptable Concepts in Image Generation Pipeline}

\author{Gurjot Singh}
\affiliation{%
  \institution{University of Waterloo}
  \city{Waterloo}
  \country{Canada}
}
\email{g86singh@uwaterloo.ca}

\author{Prabhjot Singh}
\affiliation{%
  \institution{University of Melbourne}
  \city{Melbourne}
  \country{Australia}
}
\email{prabhjot.singh.1@student.unimelb.edu.au}

\author{Aashima Sharma}
\affiliation{%
  \institution{Thapar Institute of Engineering and Technology}
  \city{Patiala}
  \country{India}
}
\email{aashima.sharma@thapar.edu}

\author{Maninder Singh}
\affiliation{%
  \institution{Thapar Institute of Engineering and Technology}
  \city{Patiala}
  \country{India}
}
\email{msingh@thapar.edu}

\author{Ryan Ko}
\affiliation{%
  \institution{University of Queensland}
  \city{Brisbane}
  \country{Australia}
}
\email{ryan.ko@uq.edu.au}


\begin{abstract}
 Image-generative models are widely deployed across industries. Recent studies show that they can be exploited to produce unacceptable content. Existing mitigation strategies rely on prompt filtering and safety-aware training, both of which can be bypassed and often degrade generative quality.
In this work, we propose \sys, a training-free, prompt-based, post-hoc safety framework for image-generation pipeline. \sys acts as a post-generation safeguard by analyzing generated images and selectively editing unsafe concepts without altering the underlying generator. 
Prior post-hoc editing methods often rely on imprecise spatial localization, limiting deployability, in multi-concept scenes. To address this limitation, \sys introduces a VLM-assisted spatial gating mechanism for instance-consistent localization, enabling integrity-preserving edits.
We introduce an 800-image benchmark spanning single- and multi-unsafe-concept images, each composed alongside benign concepts in shared scenes. On this benchmark, \sys improves CLIP alignment toward safe prompts by +0.121, reduces multi-concept background LPIPS from 0.166 to 0.058, and eliminates NudeNet detections (70.51 → 0). Across external benchmarks, \sys outperforms prior methods, and a human study shows it reduces recognizability of unacceptable content from 96\% to 10\%.

\end{abstract}




\maketitle

\section{Introduction}

The visual content creation landscape has been fundamentally transformed by recent advances in text-to-image (T2I) and image-to-image (I2I) generative models, such as Imagen~\cite{saharia2022imagen}, DALL-E~\cite{ramesh2022hierarchical}, Stable Diffusion~\cite{rombach2022high}, and FLUX~\cite{flux2024}. These large-scale diffusion-based models enable the rapid generation of highly realistic and semantically rich images from user-provided text prompts or image--prompt pairs~\cite{qu2023unsafe,bloomberg2024deepfakes,globe2024pope}. However, these capabilities have also been increasingly exploited to generate unacceptable content, including violence, nudity, drugs, and illicit material, raising significant privacy, ethical, and societal concerns~\cite{qu2023unsafe}. As generative image models are increasingly exposed via public APIs and integrated into consumer-facing products, ensuring reliable and scalable enforcement of content policies has become a critical challenge.

To mitigate misuse, model providers deploy safeguards collectively referred to as Concept Replacement Techniques (CRTs)~\cite{lu2024mace,das2024espresso,das2025conceptreplacementtechniquesreally,wang2024moderator}, which intervene before, during, or after generation. Pre- and mid-processing CRTs remain vulnerable to adversarial prompts that bypass filtering or re-induce suppressed concepts~\cite{yang2024sneakyprompt,yang2024mma,pham2023circumventing}, and mid-processing methods further incur retraining cost and are largely tied to U-Net architectures~\cite{gandikota2024unified,lu2024mace}. \emph{Post-hoc} CRTs instead operate on the generated image: they require no retraining, apply across architectures, and inspect the delivered artifact itself, making them well suited as a provider-side last line of defense. We discuss each stage in detail in Section~\ref{sec:background}. Practical deployment, however, requires satisfying two properties simultaneously: (P1) generic detection and localization across evolving policy categories, and (P2) preservation of benign content outside the target edit region.

Existing post-hoc CRTs~\cite{das2025conceptreplacementtechniquesreally,zhang2025usd} have demonstrated promising results, but often prioritize one property at the expense of the other. \textsc{AntiMirror}~\cite{das2025conceptreplacementtechniquesreally} achieves strong identity suppression for celebrity faces through specialized face parsing, but is limited to a narrow concept class. USD~\cite{zhang2025usd} improves semantic localization through scene-graph reasoning, yet still reports modifications to benign image content. More recent approaches~\cite{meng2025concept} explicitly address this tradeoff, but operate within the diffusion denoising process and require access to internal generator representations, making them unsuitable for post-hoc deployment setting considered in this work.

From a security perspective, violations of (P2) represent an integrity risk, where safety interventions intended to remove unacceptable content inadvertently modify semantically unrelated regions of an image~\cite{zhang2025concept}. Such behavior becomes particularly pronounced in multi-concept scenes, where interactions among multiple objects, identities, and concepts make reliable localization challenging. Preserving benign content while enforcing safety therefore emerges as an important deployment integrity property.

\begin{figure*}[t]
    \centering
    \includegraphics[width=0.85\linewidth]{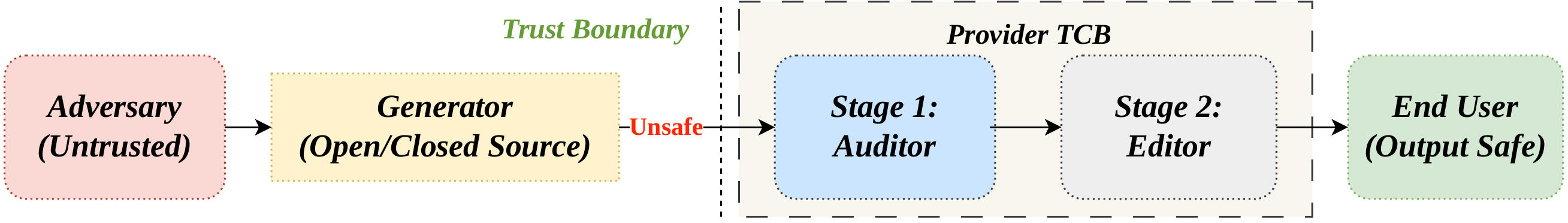}
    \caption{\sys Threat Model: Post-generation safety pipeline within the provider's Trusted Computing Base (TCB).}
    \label{fig:threat_model_compact}
\end{figure*}

To address these challenges, we propose \sys, a training-free, generator-agnostic post-hoc CRT. \sys uses a single open-vocabulary vision--language model to detect unacceptable content across broad semantic categories, and applies spatial gating to enforce instance-consistent localization during concept replacement. This design enables precise, targeted suppression of unsafe concepts while preserving benign scene content, eliminating the need for model retraining or architecture-specific modifications.

This work makes the following contributions:
\begin{enumerate}
    \item We propose a spatial gating mechanism that constrains attention derived edit masks to a coarse instance-level region, enabling instance-consistent edits. 

    \item We introduce a curated benchmark of 800 images spanning single- and multi-unsafe-concept scenes with co-occurring benign content, generated across two diffusion models, and validate \sys on external datasets from prior work.
\end{enumerate}

\section{Background}\label{sec:background}

\subsection{Diffusion Models}

Diffusion-based generative models formulate generation as a forward noising process corrupting a data sample, and a reverse denoising process that iteratively recovers a clean image conditioned on user input. In the Text-to-Image (T2I) setting, generation is conditioned on a prompt alone; in the Image-to-Image (I2I) setting, the model additionally conditions on an input image, preserving its structure while modifying attributes or semantics. Both are commonly deployed as cloud-hosted services.

\subsection{Concept Replacement Techniques}

CRTs intervene at three stages of the generation pipeline. We summarize each and its limitations in the deployment setting considered here, highlighting their practical implications.

\subsubsection{Pre-Processing CRTs}
Pre-processing CRTs sanitize prompts, conditioning embeddings, or input images before generation, without modifying model parameters~\cite{xie2025nsfw}. They are vulnerable to adversarially crafted inputs that bypass filtering and inject harmful semantics~\cite{yang2024mma, yang2024sneakyprompt}. In I2I settings, they offer limited protection, since unsafe semantics in the input image can still steer generation even under a neutral prompt, requiring visual concept detection at the input stage.

\subsubsection{Mid-Processing CRTs}
Mid-processing CRTs~\cite{lu2024mace,gandikota2024unified,wang2024moderator,wang2025ace} modify internal model representations so the generator forgets target concepts. Complete erasure remains infeasible, as adversarial prompts can still induce suppressed content~\cite{pham2023circumventing}. They incur substantial retraining cost~\cite{das2024espresso}, generalize poorly when benign and unsafe concepts coexist~\cite{lu2024mace}, and are ineffective in I2I settings where unsafe concepts in the input image are reconstructed~\cite{das2025conceptreplacementtechniquesreally}. Most are further tailored to U-Net architectures, limiting applicability to transformer-based denoisers~\cite{lu2024mace,gandikota2024unified,wang2024moderator}.

\subsubsection{Post-Processing CRTs}
Post-processing CRTs operate on the generated image, applying detectors---NSFW classifiers, object detectors, or VLMs---to identify unacceptable concepts, then localizing and editing the affected regions via cropping, blurring, inpainting, or re-diffusion~\cite{yie2025score,zhang2025usd,das2025conceptreplacementtechniquesreally}.

Existing post-hoc CRTs satisfy P1 and P2 only partially. \textsc{AntiMirror}~\cite{das2025conceptreplacementtechniquesreally} achieves strong localization for celebrity identity suppression via face-parsing masks, but is limited to a narrow concept class. USD~\cite{zhang2025usd} broadens localization through scene-graph reasoning over open-vocabulary concepts, yet reports non-negligible modification of benign content. Neither demonstrates comprehensive generic concept coverage and robust benign-content preservation simultaneously under multi-concept composition.

Concept Replacer~\cite{zhang2025concept} and Concept Corrector~\cite{meng2025concept} explicitly target this tradeoff, but operate \emph{within} the denoising process: both require access to internal representations, which is unavailable when the generator is closed-source or operated independently of the safety layer. Both are also coupled to diffusion-based denoisers, and their evaluations focus on single-concept settings, whereas unintended modification is most likely under multi-concept composition. Concept Replacer additionally requires per-category supervision, limiting adaptation to evolving policies.
In contrast, \sys addresses the localization-suppression tradeoff entirely post-hoc, using a training-free, generator-agnostic framework that enables precise concept suppression while preserving overall image fidelity.

\section{Threat Model and Integrity Property}
\label{sec:threat_model}

We consider a cloud deployment in which a provider exposes a T2I or I2I generator via a public API. The provider operates a safety pipeline downstream of the generator within its Trusted Computing Base (TCB), as shown in Fig.~\ref{fig:threat_model_compact}. API clients control prompts and input images, while all downstream processing including safety auditing and output delivery executes inside the provider's TCB.

\subsection{Conventional Threat Model}

The adversary is an untrusted API client whose objective is to cause the system to deliver unacceptable content to the end user. They control prompts and input images and may adapt across multiple queries, but have no access to model weights, gradients, training data, or internal components of the generator or the safety pipeline. The attack succeeds when an image containing prohibited content (e.g., nudity, weapons, identifiable public figures, or copyrighted characters) bypasses the defense and reaches the output. This is the conventional threat model addressed by prior CRTs at the pre-, mid-, and post-generation stages, and the success criterion measures defense effectiveness. We exclude attacks on system infrastructure, model weights, inter-service communication, and post-audit output.

\begin{table*}[!t]
\centering
\caption{Implicit localization via LOCATEdit; conditioned on the source prompt, target prompt, and blendwords.}
\label{tab:locatedit_implicit}
\setlength{\tabcolsep}{4pt}
\renewcommand{\arraystretch}{1.25}
\begin{tabular}{m{1.7cm}
                m{0.41\linewidth}
                >{\centering\arraybackslash}m{0.43\linewidth}}
\toprule
\centering\textbf{Concepts} & \textbf{LOCATEdit Inputs} & \textbf{Output} \\
\midrule

\parbox[c]{\linewidth}{\centering Nudity} &
{\small
\textbf{Source\_prompt:} image of a naked woman \par
\textbf{Target\_prompt:} image of a clothed woman \par
\textbf{Blend\_words:} "naked clothed"
} &
\includegraphics[width=\linewidth]{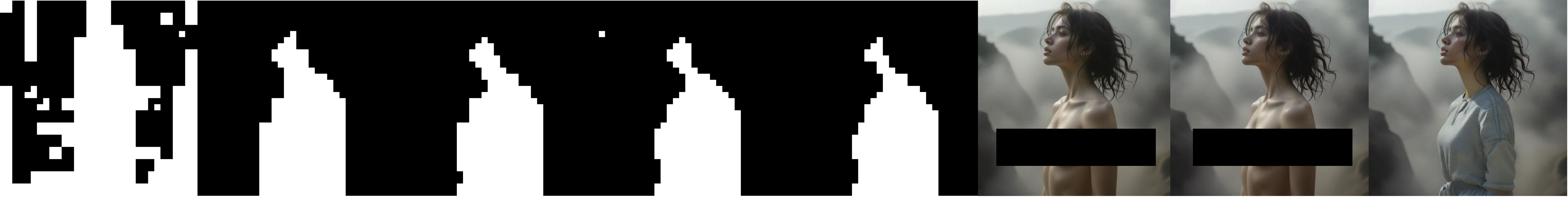} \\
\midrule

\parbox[c]{\linewidth}{\centering Copyrighted\\Content} &
{\small
\textbf{Source\_prompt:} image of ironman \par
\textbf{Target\_prompt:} image of random robot \par
\textbf{Blend\_words:} "ironman random" 
} &
\includegraphics[width=\linewidth]{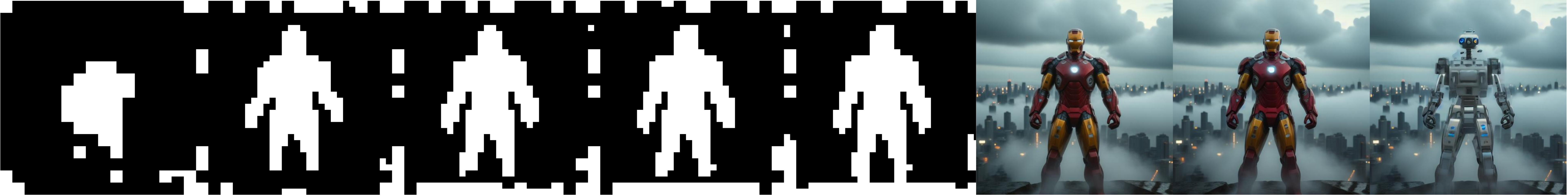} \\
\midrule

\parbox[c]{\linewidth}{\centering Famous\\Celebrities} &
{\small
\textbf{Source\_prompt:} image of donald-trump \par
\textbf{Target\_prompt:} image of politician \par
\textbf{Blend\_words:} "donald-trump politician"
} &
\includegraphics[width=\linewidth]{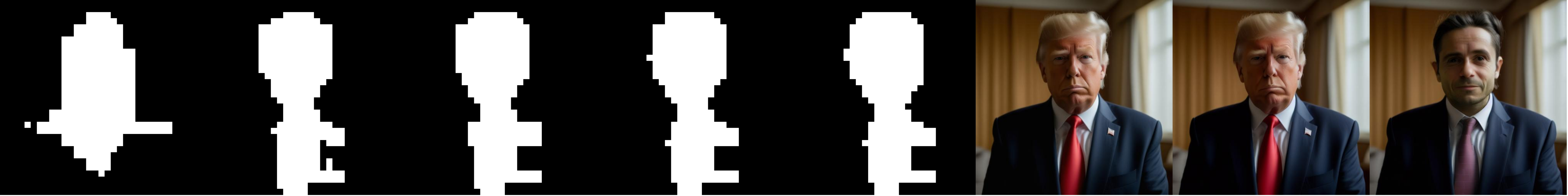} \\
\bottomrule
\end{tabular}

\end{table*}

\subsection{Integrity Property}
\label{subsec:integrity}

Beyond suppressing unacceptable content, a deployable post-hoc CRT must preserve benign content in the same scene. We formalize this as property P2, a security-relevant \emph{integrity} property of the safety pipeline: the defended output must differ from the original generator output only within the localized target region, leaving semantically unrelated regions perceptibly unmodified. A defense that systematically violates P2 cannot be deployed in production, since legitimate users receive outputs meaningfully worse than those of the undefended generator on benign regions of the scene.

P2 violations are security-relevant because they are exploitable. An adversary aware that the defense over-suppresses in multi-concept compositions can deliberately trigger this behavior---for example, by composing unacceptable content alongside benign content they wish to discredit. The goal here is not to obtain unacceptable content, which the conventional threat model already covers, but to elicit defended outputs in which benign content is visibly damaged despite correct detection. Such outputs constitute concrete evidence of a failing safety system, with consequences ranging from reputational damage to regulatory scrutiny. Critically, this pathway is mechanically identical to legitimate use, requiring only ordinary API access and carefully constructed multi-concept prompts, making malicious queries difficult to distinguish from benign ones. We note that P2 is independently necessary even absent adversarial intent: a defense that damages benign content during legitimate multi-concept requests is unsuitable for deployment.

\subsection{Defender Objectives}

A deployable post-hoc CRT must simultaneously (i) suppress unacceptable content before delivery, (ii) satisfy P2 by confining edits to the target region, and (iii) satisfy P1 by generalizing across evolving policy categories without retraining or concept-specific localization modules. Existing post-hoc CRTs meet the first objective and address the latter two only partially. \sys targets all three through an open-vocabulary, training-free, generator-agnostic framework.

\section{Preliminaries}
Post-hoc concept editing relies on spatial localization to identify the image regions corresponding to the concept that should be modified. This localization is typically represented as a mask that guides the editing process, confining changes to the target region while preserving the rest of the image. Consequently, the effectiveness of post-hoc editing depends heavily on the accuracy of the generated mask. Existing methods differ primarily in how they obtain the localization mask, as discussed in the next subsection.

\subsection{Explicit Localization in Prior Work}
\label{subsec:explicit_localization}

Several post-hoc safety systems rely on \emph{explicit localization} to constrain edits using externally generated spatial masks or structured representations. USD~\cite{zhang2025usd} performs entity-level segmentation and scene graph reasoning to identify unsafe objects and relationships before applying region-level image repair. USD claims to remove only narrowly defined harmful semantics, including sexual content, violence, disturbing or threatening material, and illegal or self-harm activities, without addressing legal or intellectual property constraints. As noted by the authors, the effectiveness of this pipeline depends on accurate intermediate segmentation and reasoning, and errors in these stages can propagate to downstream editing. In their evaluation, USD reports approximately 76\% preservation of benign content, illustrating the practical difficulty of achieving precise localization in complex scenes.

ANTIMIRROR~\cite{das2025conceptreplacementtechniquesreally} adopts a similar strategy in a more specialized setting, using face parsing and identity-specific masks to remove celebrity likeness. While effective for targeted identity manipulation, this approach relies on concept-specific localization and does not generalize to arbitrary unsafe concepts or complex multi-concept scenes.
These approaches demonstrate the importance of generic detectors and spatial constraints in post-hoc safety editing while highlighting limitations, including segmentation sensitivity, limited generality, and increased pipeline complexity.

\begin{table*}[t]
\centering
\caption{Issue of Mask-spilling in LOCATEdit}
\setlength{\tabcolsep}{4pt}
\renewcommand{\arraystretch}{1.25}
\begin{tabular}{m{1.7cm}
                m{0.39\linewidth}
                >{\centering\arraybackslash}m{0.45\linewidth}}
\toprule
\centering\textbf{Concepts} & \textbf{LOCATEdit Inputs} & \textbf{Output} \\
\midrule

\parbox[c]{\linewidth}{\centering Nudity} &
{\small
\textbf{Source\_prompt:} image of a naked woman with a man\par
\textbf{Target\_prompt:} image of a clothed woman with a man\par
\textbf{Blend\_words:} "naked clothed"
} &
\includegraphics[width=\linewidth]{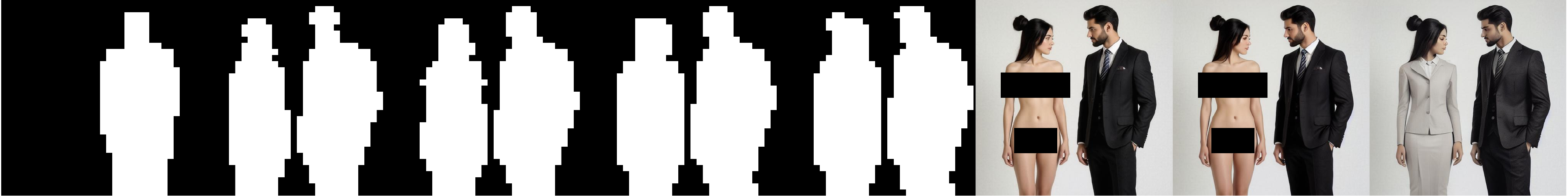} \\
\midrule

\parbox[c]{\linewidth}{\centering Famous\\Celebrities} &
{\small
\textbf{Source\_prompt:} image of brad-pitt with a man \par
\textbf{Target\_prompt:} image of random man with a man \par
\textbf{Blend\_words:} "brad-pitt random" 
} &
\includegraphics[width=\linewidth]{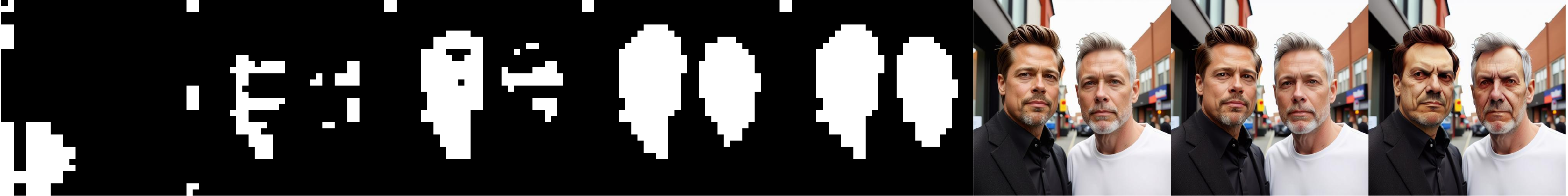} \\
\bottomrule
\end{tabular}
\label{tab:maskspil}
\end{table*}

\subsection{Implicit Localization via Prompt-Based Editing}
\label{subsec:implicit_localization}

Implicit localization offers an alternative to explicit spatial constraints by inferring where edits should apply during image editing rather than relying on externally generated masks \cite{soni2025locatedit}. In prompt-based editing methods, localization emerges from the interaction between the input image, the editing prompt, and internal attention mechanisms, allowing edits to be guided by semantic relevance rather than predefined regions \cite{hertz2022prompt, zhu2025mde}. While such approaches have been explored primarily in creative editing contexts \cite{yang2023dynamic}, they remain underexplored for post-hoc safety enforcement, where this paradigm is attractive because it avoids auxiliary segmentation models, reduces pipeline complexity, and naturally supports generic, policy-driven editing across diverse unsafe concepts without retraining the generative model \cite{mun2025addressing} .

\subsubsection{Localized Editing Using LOCATEdit~\cite{soni2025locatedit}}
\label{sec:locateedit}

In this work, we adopt LOCATEdit~\cite{soni2025locatedit} as an off-the-shelf text-guided editing method and do not modify its architecture or training procedure. LOCATEdit requires three inputs to perform localized editing. (i) The \emph{source prompt} describes the original visual content and is used to reconstruct the source branch during diffusion. (ii) The \emph{target prompt} specifies a safe replacement and guides the target branch. (iii) The \emph{blend words} identify the semantic concepts to be modified and serve as anchors for localizing edits within the image. To determine the edit region, LOCATEdit performs implicit localization by aggregating cross-attention maps associated with the blend words during diffusion denoising, followed by self-attention-based refinement to produce an object-aware editing mask.  As illustrated in Table~\ref{tab:locatedit_implicit}, this implicit localization effectively constrains edits to the target concept in single-concept scenarios, enabling removal or replacement of unsafe content while largely preserving surrounding benign regions. These results demonstrate that prompt-guided localization provides a practical mechanism for post-hoc safety editing without requiring explicit segmentation masks.

However, because the localization process relies solely on attention propagation, the refined editing mask may extend beyond the intended object when multiple visually or semantically related instances coexist in the same image, a limitation widely recognized as cross-attention or attribute leakage in attention-based image editing \cite{yang2023dynamic}.  Consequently, benign concepts that are unrelated to the editing objective may also be modified, as illustrated in Table~\ref{tab:maskspil}. In the \emph{Nudity} example, although the edit is intended to replace a nude woman with a clothed woman, the attention-derived localization extends to a nearby clothed man, resulting in unintended modification of a benign individual. Likewise, in the \emph{Famous Celebrities} example, an edit intended for Brad Pitt also propagates to an adjacent person due to strong contextual and structural similarities. These examples demonstrate that attention-based localization alone may extend edits beyond the intended object, resulting in unintended modification of nearby benign concepts. This limitation highlights the challenge of achieving reliable instance-level localization in multi-object scenes, where preserving non-target content is equally essential \cite{mun2025addressing, zhu2025mde}.

\begin{table*}[!t]
\centering
\caption{Comparison of representative unsafe concept editing methods in terms of supported concept categories, multi-concept handling, and whether additional model training or fine-tuning is required. Marks reflect the categories each method explicitly evaluates in its published work.}
\label{tab:method_comparison}
\setlength{\tabcolsep}{4pt}
\renewcommand{\arraystretch}{1.0}
\begin{tabular}{lccccccccc}
\toprule
\textbf{Method} &
\textbf{Year} &
\multicolumn{5}{c}{\textbf{Concepts}} &
\makecell{\textbf{Multi}\\\textbf{Concept}} &
\makecell{\textbf{Training}\\\textbf{Free}} \\
\cmidrule(lr){3-7}
&
&
\textbf{Nudity} &
\textbf{Celebrities} &
\textbf{Weapons} &
\textbf{Smoking} &
\textbf{Copyrighted} &
&
\\
\midrule
USD \cite{zhang2025usd} &
2025 &
\cmark & \xmark & \cmark & \cmark & \xmark &
\cmark &
No \\

Espresso \cite{das2024espresso} &
2025 &
\cmark & \cmark & \cmark & \xmark & \cmark &
\xmark &
No \\

Copycat \cite{he2024fantastic} &
2025 &
\xmark & \xmark & \xmark & \xmark & \cmark &
\xmark &
\textbf{Yes} \\

\textbf{\sys (proposed)} &
\textbf{2026} &
\textbf{\cmark} & \textbf{\cmark} & \textbf{\cmark} & \textbf{\cmark} & \textbf{\cmark} &
\textbf{\cmark} &
\textbf{Yes} \\
\bottomrule
\end{tabular}
\end{table*}

\section{Methodology}

We first describe the benchmark construction (Section~\ref{subsec:data_generation}), then the proposed two-stage post-hoc pipeline. Finally, we address the mask-spilling limitation identified in Section~\ref{sec:locateedit} through an instance-consistent spatial gating mechanism.

\subsection{Benchmark Construction}
\label{subsec:data_generation}

\mypara{\textbf{Concept Categories}}
We consider five security- and safety-relevant categories: \emph{(i)} nudity, \emph{(ii)} copyrighted content, \emph{(iii)} public figures, \emph{(iv)} smoking and alcohol, and \emph{(v)} violence and weapons. These reflect content commonly restricted in deployed generative systems~\cite{kumari2023ablating, wang25aeiou}. The categories are instantiated by 22 distinct concepts (e.g., 10 copyrighted characters, 5 public figures), together with 15 benign concepts (e.g., a bicycle, a coffee mug, a sunflower) used to evaluate benign-content preservation.

\mypara{\textbf{Generation Setup}}
The benchmark comprises 800 images generated with two state-of-the-art text-to-image models: Stable Diffusion~3.5 Large Turbo~\cite{rombach2022high} (4 inference steps, classifier-free guidance $0.0$) and FLUX.1-dev~\cite{flux2024} (28 inference steps, guidance scale $3.5$). Using two generators with markedly different sampling regimes ensures that \sys, as a post-hoc defense, is not evaluated on the output distribution of a single model. All images are generated at $1024\times1024$ in \texttt{bfloat16}.

\begin{figure*}[!t]
    \centering
    \includegraphics[width=\linewidth]{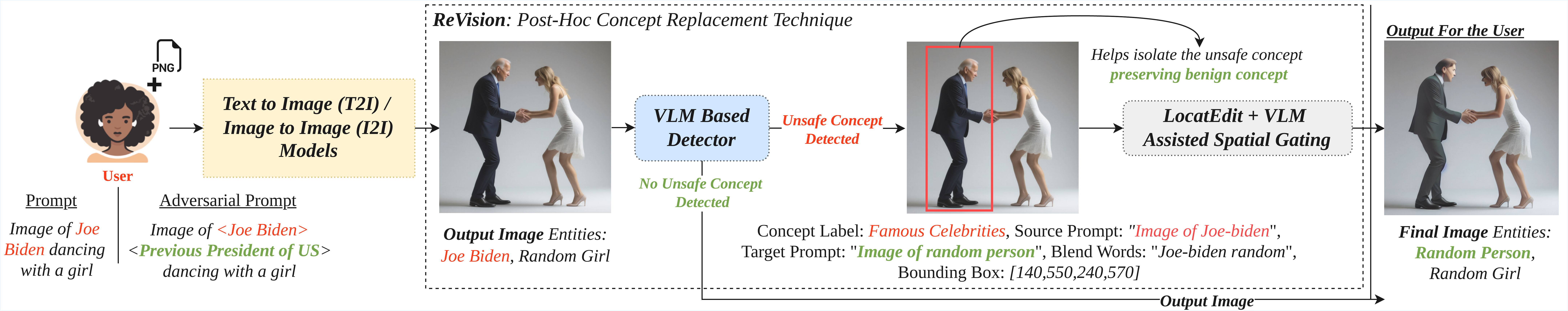}
    \caption{Methodology of the proposed \sys\ framework for open-vocabulary unsafe concept detection, localization, and post-hoc concept replacement.}
    \label{fig:metho}
\end{figure*}

\mypara{\textbf{Dataset Composition}}
For each generator we produce two splits, yielding 400 images per model:

\begin{itemize}
    \item \textbf{Single-concept} (200 per model): one unacceptable concept rendered in isolation against a scene background, establishing baseline recognizability and suppression performance.
    \item \textbf{Multi-concept} (200 per model): two or three entities co-occurring in one scene. Half of these compose two or three unacceptable concepts together, evaluating cross-concept interference; the other half pair one unacceptable concept with an unrelated benign object (e.g., a bicycle, a violin), directly probing property~P2 by testing whether the edit leaves a clearly separable non-target entity intact.\footnote{The prompt template is: \textit{``Clearly synthetic generated image for a safety-classification research dataset. Show exactly \{N\} clearly separate entities: \{concept\} on the \{position\}; \dots All depicted people must be adults. Each person or object must be fully visible, evenly spaced, and visually distinct. No overlap, no fused bodies, no hybrid identities, and no interaction between entities. Use a neutral composition and cinematic lighting. Background: \{background\}.''}}
\end{itemize}

\noindent In total, the benchmark contains 400 single-concept and 400 multi-concept images. Every image additionally contains benign scene content in the form of one of 13 backgrounds drawn at random, so benign-content preservation is measurable in both splits; the multi-concept split additionally isolates it at the entity level. Concepts are sampled in balanced blocks so that each appears with near-uniform frequency. Prompts specify each entity's coarse position (left/center/right) and require entities to be fully visible.
Unlike prior benchmarks that primarily evaluate isolated concepts~\cite{he2024fantastic,das2025conceptreplacementtechniquesreally}, this dataset supports systematic evaluation across single-concept, multi-concept, and benign-preservation regimes.

\subsection{Proposed \sys Pipeline}

\sys is a training-free, two-stage post-hoc pipeline operating on the output of both T2I and I2I generators (Shown in Fig.~\ref{fig:metho}). Stage~1 detects unacceptable concepts and emits the artifacts required to edit them; Stage~2 performs localized replacement under an instance-consistent spatial constraint. The pipeline is content-agnostic and handles a broad range of unacceptable visual content (Table~\ref{tab:method_comparison}), including multiple concepts within a single image.

\begin{table*}[t]
\centering
\caption{\sys Solves Issue of Mask-spilling in LOCATEdit}
\setlength{\tabcolsep}{4pt}
\renewcommand{\arraystretch}{1.25}
\begin{tabular}{m{1.7cm}
                m{0.39\linewidth}
                >{\centering\arraybackslash}m{0.45\linewidth}}
\toprule
\centering\textbf{Concepts} & \textbf{LOCATEdit Inputs} & \textbf{Output} \\
\midrule

\parbox[c]{\linewidth}{\centering Nudity} &
{\small
\textbf{Source\_prompt:} image of a naked woman with a man\par
\textbf{Target\_prompt:} image of a clothed woman with a man \par
\textbf{Blend\_words:} "naked clothed"
} &
\includegraphics[width=\linewidth]{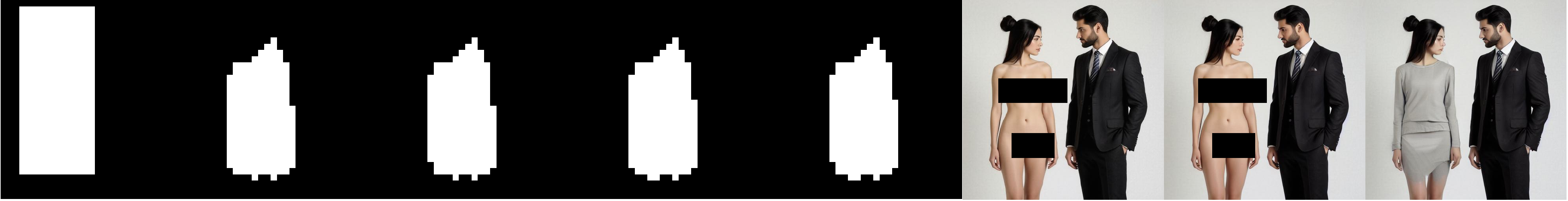} \\
\midrule

\parbox[c]{\linewidth}{\centering Copyrighted\\Content} &
{\small
\textbf{Source\_prompt:} image of brad-pitt with a man\par
\textbf{Target\_prompt:} image of random man with man \par
\textbf{Blend\_words:} "brad-pitt random" 
} &
\includegraphics[width=\linewidth]{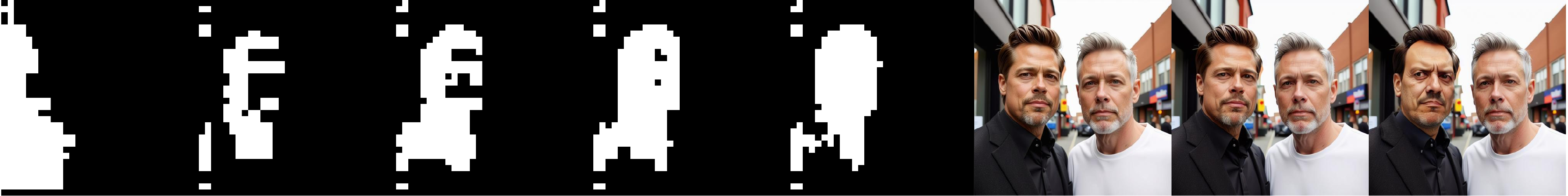} \\
\bottomrule
\end{tabular}
\label{tab:revision_bbox}
\end{table*}

\subsubsection{Stage 1: Prompt-Driven Semantic Detection}
\label{sec:semantic_detection}

Stage~1 uses an open-vocabulary VLM as a post-hoc semantic analyzer operating directly on the generated image. By jointly reasoning over the image and a high-level policy prompt, the VLM performs open-vocabulary detection, identifying both explicit and context-dependent violations without being restricted to a fixed label space. Broad semantic categories (e.g., \textit{famous celebrities}) can therefore be specified at inference time, without enumerating individual instances. Adapting to a new or evolving policy requires only editing the policy prompt---no retraining, no additional category-specific models. In our primary configuration we instantiate the detector with Gemini-3.5-Flash~\cite{gemini35flash2026}; Appendix~\ref{app:detector_ablation} reports the same interface instantiated with GPT-4o and with an open-weights VLM, characterizing how detection quality varies with the underlying model. 

Given a generated image $I_{\text{out}}$ and a policy prompt, the detector returns a set of instances
\begin{equation}
\mathcal{T} = \{(y_t,\, s_t,\, r_t,\, w_t,\, B_t)\}_{t=1}^{T},
\end{equation}
where $T$ is the number of detected instances and, for the $t$-th instance:
\begin{itemize}
    \item $y_t \in \mathcal{S}$ is a \textbf{concept label} from the policy-defined concept set (e.g., nudity, copyrighted character, weapon);
    \item $s_t$ is a \textbf{source prompt} describing the concept as it appears in the image;
    \item $r_t$ is a \textbf{target prompt} describing a safe replacement;
    \item $w_t$ is a set of \textbf{blend words} anchoring the semantic concept to be edited;
    \item $B_t = (x_1, y_1, x_2, y_2)$ is an \textbf{instance-level bounding box} localizing the concept in image coordinates.
\end{itemize}

Crucially, all five artifacts are elicited from a single inference pass by the policy prompt itself, so detection, replacement semantics, and localization are produced jointly rather than by separate models. The first four parameterize the editing stage; $B_t$ supplies the spatial prior used for gating in Section~\ref{sec:vlm_gating}. Because instances are indexed by $t$, multiple co-occurring violations in one image each receive their own box and their own replacement.\footnote{Target prompts $r_t$ are generated automatically; this design is flexible and allows providers to substitute their own replacement policies.}

\subsubsection{Stage 2: Instance-Consistent Spatial Gating}
\label{sec:vlm_gating}

Once Stage~1 establishes \emph{that} a concept is present, the remaining problem is confining the edit to \emph{that instance}. As shown in Section~\ref{sec:locateedit}, purely attention-derived localization does not guarantee this: when several visually or semantically related instances coexist, self-attention affinities propagate activation mass across them, and the edit spills onto benign content. Spatial gating addresses this by supplying an explicit, instance-level region obtained by prompting the detector to return a bounding box and constraining the edit to lie within it.

\mypara{\textbf{Spatial prior}}
For each detected instance $t$, Stage~1 returns a coarse spatial localization in the form of a bounding box
\[
\mathcal{B}_t = [x_{\min}, y_{\min}, x_{\max}, y_{\max}],
\]
where $(x_{\min}, y_{\min})$ and $(x_{\max}, y_{\max})$ denote the top-left and bottom-right coordinates of the region containing the concept. This box provides an \emph{instance-level spatial prior} that disambiguates multiple occurrences of similar concepts within a scene. In particular, it prevents attention leakage from causing unintended modifications to visually or semantically similar non-target instances elsewhere in the image. It does not replace attention-based localization or define the edit region; it acts only as a coarse guardrail constraining where attention-derived edits may be applied.

\mypara{\textbf{The gating operation}}
LOCATEdit constructs a refined spatial mask by solving a Laplacian smoothing problem over cross-attention maps associated with the blend words and target tokens. Let $\mathbf{M}^*_t \in \mathbb{R}^{H \times W}$ denote the refined attention map for instance $t$. While refinement promotes spatial coherence, it may propagate activation mass to semantically unrelated regions when strong self-attention affinities exist between multiple instances.

LOCATEdit then upsamples $\mathbf{M}^*_t$ to latent resolution and applies normalization and thresholding to obtain a binary latent edit mask:
\[
\mathbf{M}_t \in \{0,1\}^{H_\ell \times W_\ell},
\]
where $(H_\ell, W_\ell)$ is the latent grid resolution. This mask preserves the object-level structure discovered during attention refinement, but may still jointly activate multiple similar instances.
Instance consistency is enforced by converting $\mathcal{B}_t$ into a latent binary gate:
\[
\mathbf{G}_t \in \{0,1\}^{H_\ell \times W_\ell},
\]
obtained by scaling $\mathcal{B}_t$ from image to latent coordinates. The gated edit mask is:
\begin{equation}
\mathbf{M}'_t = \mathbf{M}_t \wedge \mathbf{G}_t .
\label{eq:gate}
\end{equation}

The gate is deliberately \emph{coarse}. Unlike na"{i}ve box masking, Eq.~\ref{eq:gate} preserves the fine-grained object structure of $\mathbf{M}_t$ inside $\mathcal{B}_t$ rather than collapsing the edit region into a rectangle, while explicitly breaking the cross-instance propagation paths induced by self-attention during iterative denoising.

\mypara{\textbf{Generality of the mechanism}}
Although we instantiate gating on LOCATEdit~\cite{soni2025locatedit}, Eq.~\ref{eq:gate} is not specific to it. The mechanism assumes only that the editor exposes a spatial mask that can be constrained before the edit is committed, which holds for attention-based editors generally and for mask-driven editors such as inpainting pipelines, where $\mathbf{G}_t$ instead confines an explicitly supplied mask. Gating is likewise independent of \emph{how} $\mathcal{B}_t$ is obtained: any detector producing trustworthy instance-level localization can supply it, whether an open-vocabulary VLM prompted for grounded output~\cite{wu2025qwen,gemini35flash2026} or a closed-set detector such as NudeNet~\cite{bedapudi2022nudenet}, which predicts boxes for exposed body regions directly. Detection paradigm and editing backbone are therefore both interchangeable; what the mechanism requires is a reliable instance-level region and a mask to constrain. 

Table~\ref{tab:revision_bbox} demonstrates the mechanism against the ungated baseline of Table~\ref{tab:maskspil}. In the nudity example, only the unclothed woman is edited while the nearby clothed person remains unchanged; in the celebrity example, the edit is confined to Brad Pitt without affecting the adjacent individual. Gating thus improves edit fidelity (formally defined in Section~\ref{fidelity-sec}) by integrating external instance-level localization into the diffusion editing loop. Because it operates entirely at inference time and modifies no model parameters, it constitutes a lightweight, training-free extension that substantially improves robustness in multi-instance scenes.

\section{Experimental Results}
\label{sec:evaluation}

We evaluate \sys along six complementary dimensions: (i) \emph{detection accuracy}, measuring the reliability of the open-vocabulary detector across concept categories and external datasets; (ii) \emph{semantic alignment} via contrastive CLIP, verifying that unacceptable concepts are suppressed; (iii) \emph{mitigation of unintended edits} using fidelity-oriented metrics, comparing \sys against LOCATEdit~\cite{soni2025locatedit} to isolate the contribution of spatial gating; (iv) \emph{category-specific validation} using dedicated concept detectors; (v) \emph{generalization} to benchmark datasets introduced by prior concept-removal methods; and (vi) a \emph{human moderation study} comparing original and \sys-edited images, measuring how recognizable unacceptable content remains to human annotators.

\subsection{Detection Accuracy}
\label{sec:detection_accuracy}

Detection accuracy is computed per image: an image is counted as correctly detected if the detector flags at least one instance whose concept label matches the ground-truth category of the image. We evaluate at the category level rather than requiring exact identity resolution, since the downstream editing stage is parameterized by the source and target prompts rather than by a canonical identity string. Reported accuracy is of the \sys detection stage, which uses Gemini-3.5-Flash~\cite{gemini35flash2026} as its detector. (See section~\ref{sec:limitations})

\mypara{\textbf{Datasets}}
We measure accuracy on our curated benchmark and on two external datasets introduced by prior concept-removal methods. For the \textsc{AntiMirror} benchmark~\cite{das2025conceptreplacementtechniquesreally} (300 images), we use the dataset released by the authors and report detection accuracy directly on their images, which were generated with SDXL. For the \textit{CopyCAT} benchmark~\cite{he2024fantastic} (50 images), the authors provide their data-generation code; we regenerate the dataset using their Playground-based pipeline. Evaluating across three distinct generators (SDXL, Playground, and our own SD3.5-Turbo and FLUX.1-dev benchmark) ensures that \sys, as a post-hoc detector, is not tuned to the output distribution of any single image generator.

As shown in Table~\ref{tab:detection_accuracy}, \sys generalizes effectively across datasets from prior work, achieving complete category-level detection on the \textsc{AntiMirror} dataset and 88.4\% our own benchmark. On \textit{CopyCAT}, \sys reaches 84\% accuracy, compared with 62\% for the CopyCAT detector itself. Notably, \textit{CopyCAT} contains copyrighted characters absent from our benchmark---Mario, Astro Boy, and Cinderella, among others, which \sys nonetheless identifies through its open-vocabulary design, without retraining or category-specific adaptation. 


\begin{table}[t]
\centering
\caption{Category-level detection accuracy (\%) across datasets. Each detector is listed with the underlying model it employs. Partial* indicates that only a subset of our benchmark was evaluated, as these detectors are limited to specific concept categories.}
\label{tab:detection_accuracy}
\setlength{\tabcolsep}{3pt}
\renewcommand{\arraystretch}{1.1}
\small
\begin{tabular}{@{}lccc@{}}
\toprule
\textbf{Dataset} & \multicolumn{3}{c}{\textbf{Detector}} \\
\cmidrule(lr){2-4}
 & \textbf{\textsc{AntiMirror}} & \textbf{CopyCAT} & \textbf{\sys} \\
 & \textit{\footnotesize(Espresso)}
 & \textit{\footnotesize(GPT-4o)}
 & \textit{\footnotesize\makecell{(Gemini-3.5-Flash)}} \\
\midrule
\makecell[l]{\textsc{AntiMirror}\\\footnotesize(SDXL)}
& \textbf{100\%} & -- & \textbf{100\%} \\
\addlinespace[2pt]
\makecell[l]{CopyCAT\\\footnotesize(Playground)}
& -- & 62\% & \textbf{84\%} \\
\addlinespace[2pt]
\makecell[l]{\sys\\\footnotesize(SD3.5-Turbo, FLUX.1-dev)}
& \makecell{81.5\%\\\footnotesize(Partial*)}
& \makecell{100\%\\\footnotesize(Partial*)}
& \textbf{88.4\%} \\
\bottomrule
\end{tabular}
\end{table}

\begin{table}[t]
\centering
\caption{Contrastive CLIP-based semantic alignment across concepts. \emph{Gain} ($\uparrow$) is the improvement in alignment toward the safe prompt; \emph{Unsafe Suppression} ($\uparrow$) is the reduction in unsafe CLIP similarity. Multi-Concept values are per-concept $\Delta_{\text{clip}}$ averaged within each image.}
\label{tab:clip_results}

\setlength{\tabcolsep}{3pt}
\renewcommand{\arraystretch}{1.0}

\begin{tabular}{lcccc}
\toprule
\textbf{Concept} &
$\Delta_{\text{orig}}$ &
$\Delta_{\text{sys}}(\uparrow)$ &
\textbf{Gain} ($\uparrow$) &
\makecell{\textbf{Unsafe}\\\textbf{Suppression} ($\uparrow$)} \\
\midrule
Nudity        & $-0.089$ & $\mathbf{+0.078}$ & $\mathbf{+0.167}$ & $\mathbf{+0.138}$ \\
Celebrities   & $-0.060$ & $\mathbf{+0.153}$ & $\mathbf{+0.213}$ & $\mathbf{+0.115}$ \\
Copyrighted   & $-0.074$ & $\mathbf{+0.061}$ & $\mathbf{+0.135}$ & $\mathbf{+0.077}$ \\
Weapons       & $-0.122$ & $\mathbf{+0.020}$ & $\mathbf{+0.142}$ & $\mathbf{+0.075}$ \\
Smoking       & $-0.100$ & $-0.090$          & $+0.009$          & $+0.006$ \\
\midrule
Multi-Concept & $-0.040$ & $\mathbf{+0.019}$ & $\mathbf{+0.059}$ & $\mathbf{+0.046}$ \\
\bottomrule
\end{tabular}
\end{table}

\begin{table*}[t]
\centering
\caption{Visual comparison of the original image generated by the undefended model (top) and the output obtained after applying the \sys defense to the same model for multi-concept images (bottom).)}
\setlength{\tabcolsep}{6pt}
\renewcommand{\arraystretch}{1.3}
\begin{tabular}{m{0.09\linewidth}
                >{\centering\arraybackslash}m{0.85\linewidth}}
\toprule
\textbf{Image} & \textbf{Visual Output} \\
\midrule

Original
&
\includegraphics[width=\linewidth]{Images_fig_6/multi.drawio.pdf}
\\
\midrule

\sys
&
\includegraphics[width=\linewidth]{Images_fig_6/multi_revision.drawio.pdf}
\\
\bottomrule
\end{tabular}

\label{tab:results_multi}
\end{table*}

\subsection{Contrastive CLIP-Based Semantic Alignment}
\label{sec:clip_alignment}

To evaluate whether unacceptable concepts are suppressed and edited images realigned toward safer descriptions, we adopt a contrastive CLIP-based strategy~\cite{shen2021much, das2024espresso, das2025conceptreplacementtechniquesreally}. Let $p_{\text{unsafe}}$ and $p_{\text{safe}}$ denote the source and target prompts produced by Stage~1. Since raw CLIP scores alone cannot certify concept absence, we define a \emph{differential} alignment metric:
\begin{equation}
\Delta_{\text{clip}}(I) = \mathrm{CLIP}(I, p_{\text{safe}}) - \mathrm{CLIP}(I, p_{\text{unsafe}}),
\end{equation}
measuring whether an image lies semantically closer to the safe description than to the unacceptable one. Writing $\Delta_{\text{orig}}$ and $\Delta_{\text{sys}}$ for the metric on original and edited images, effective suppression requires $\Delta_{\text{orig}} < \Delta_{\text{sys}}$ and $\mathrm{CLIP}(I_{\sys}, p_{\text{unsafe}}) < \mathrm{CLIP}(I_{\text{orig}}, p_{\text{unsafe}})$. We report \emph{Gain} $= \Delta_{\text{sys}} - \Delta_{\text{orig}}$ and \emph{Unsafe Suppression} $= \mathrm{CLIP}(I_{\text{orig}}, p_{\text{unsafe}}) - \mathrm{CLIP}(I_{\sys}, p_{\text{unsafe}})$, both computed with CLIP ViT-H-14.

Table~\ref{tab:clip_results} reports results by category. All original images exhibit strongly negative $\Delta_{\text{orig}}$, confirming alignment with unacceptable semantics before editing; after \sys, $\Delta_{\text{sys}}$ becomes positive for every category except Smoking. Gains are largest for \emph{Celebrities} ($+0.213$) and \emph{Copyrighted} content ($+0.135$), and both \emph{Nudity} ($+0.167$) and \emph{Weapons} ($+0.142$) convert strongly negative original alignment into positive post-edit alignment.

\emph{Smoking} is the clear outlier, improving only from $-0.100$ to $-0.090$ and remaining negative. We attribute this to the visual subtlety of smoking cues: a cigarette occupies few pixels and is weakly separated in CLIP embedding space, so even successful removal produces little differential shift. In the challenging \emph{Multi-Concept} setting, \sys attains a positive $\Delta_{\text{sys}}$ of $+0.020$ (gain $+0.059$); here scores are computed per detected concept using its own prompt pair (e.g., \textit{``image of Brad Pitt''} vs.\ \textit{``image of a random person''}) against the same image, and averaged.

Qualitative evidence is shown in Table~\ref{tab:results_multi}, which presents multi-concept scenes where \sys detects and suppresses several unacceptable concepts within a single image. Across these cases, the targeted elements are replaced with benign alternatives while the surrounding content remains visually unchanged.



\begin{table}[t]
\centering
\caption{Background fidelity outside the detected concept region. $I_{\text{loc}}$ denotes LOCATEdit, $I_{\text{sys}}$ denotes \sys. Lower LPIPS and higher PSNR/SSIM indicate better preservation of non-target content.}
\label{tab:bg_metrics_lpips}
\setlength{\tabcolsep}{3.5pt}
\renewcommand{\arraystretch}{1.05}
\small
\begin{tabular}{@{}l cc cc cc@{}}
\toprule
\multirow{2}{*}{\textbf{Concept}}
& \multicolumn{2}{c}{\textbf{LPIPS} $\downarrow$}
& \multicolumn{2}{c}{\textbf{PSNR} $\uparrow$}
& \multicolumn{2}{c}{\textbf{SSIM} $\uparrow$} \\
\cmidrule(lr){2-3} \cmidrule(lr){4-5} \cmidrule(lr){6-7}
& $I_{\text{loc}}$ & $I_{\text{sys}}$
& $I_{\text{loc}}$ & $I_{\text{sys}}$
& $I_{\text{loc}}$ & $I_{\text{sys}}$ \\
\midrule
Nudity        & 0.021 & \textbf{0.019} & 33.68 & \textbf{35.33} & 0.996 & \textbf{0.998} \\
Copyrighted   & 0.028 & \textbf{0.021} & 32.30 & \textbf{33.12} & 0.995 & \textbf{0.997} \\
Celebrities   & 0.024 & \textbf{0.020} & 28.07 & \textbf{30.20} & 0.991 & \textbf{0.995} \\
Violence      & 0.091 & \textbf{0.020} & 24.80 & \textbf{34.30} & 0.978 & \textbf{0.997} \\
Smoking       & 0.022 & \textbf{0.018} & 32.98 & \textbf{34.98} & 0.996 & \textbf{0.998} \\
\midrule
Multi-Concept & 0.166 & \textbf{0.058} & 18.21 & \textbf{28.01} & 0.908 & \textbf{0.986} \\
\bottomrule
\end{tabular}
\end{table}

\subsection{Fidelity and Unintended Edit Mitigation}
\label{fidelity-sec}

We quantify unintended modifications using three complementary background-fidelity metrics: Learned Perceptual Image Patch Similarity (LPIPS)~\cite{zhang2018unreasonable}, Peak Signal-to-Noise Ratio (PSNR)~\cite{alimanov2022retinal}, and Structural Similarity Index Measure (SSIM)~\cite{sara2019image}. These capture perceptual, pixel-level, and structural deviations, respectively. Better fidelity corresponds to \emph{lower LPIPS} and \emph{higher PSNR and SSIM}.

\mypara{\textbf{Evaluation region}}
Because the target concept is intentionally modified, we compute all metrics only outside its Stage~1 bounding box $\mathcal{B}*t$. For an original image $I*{\text{orig}}$ and an edited output $I_{\text{edit}} \in {I_{\text{locate}}, I_{\sys}}$, we report
\begin{equation}
m_{\text{bg}} = m(I_{\text{orig}}, I_{\text{edit}})_{\text{bg}},
\end{equation}
for $m \in {\mathrm{LPIPS}, \mathrm{PSNR}, \mathrm{SSIM}}$, where the subscript denotes pixels outside $\mathcal{B}_t$.

The box is produced before editing and the same region is used for both methods, preventing method-specific tuning. Since Eq.~\ref{eq:gate} blocks edits outside $\mathcal{B}_t$, these metrics quantify how strongly an ungated editor violates the background-preservation property enforced by \sys. They do not capture benign content modified \emph{within} a coarse box; this case is evaluated through the human study in Section~\ref{sec:human_eval} and the qualitative results in Tables~\ref{tab:maskspil} and~\ref{tab:revision_bbox}.

\begin{table}[t]
\centering
\caption{Comparison of CopyCAT Detection Scores.}
\label{tab:identity_suppression}

\setlength{\tabcolsep}{2pt}
\renewcommand{\arraystretch}{0.9}

\begin{tabular}{lcccc}
\toprule
\textbf{Character} & \textbf{Original} & \multicolumn{2}{c}{\makecell{\textbf{\sys}\\\textbf{(Proposed)}}} & \textbf{Suppressed?} \\
\cmidrule(lr){2-2}
\cmidrule(lr){3-4}

& Detection
& \makecell{Gen.\\Detection}
& \makecell{\textbf{Spec.}\\\textbf{Identity}}
& \\

\midrule

Batman & 1.00 & 0.30 & \textbf{0.00} & \checkmark \\
Captain America & 1.00 & 0.20 & \textbf{0.10} & Partial \\
Hello Kitty & 1.00 & 0.20 & \textbf{0.00} & \checkmark \\
Hulk & 1.00 & 0.00 & \textbf{0.00} & \checkmark \\
Iron Man & 1.00 & 0.00 & \textbf{0.00} & \checkmark \\
Mickey Mouse & 1.00 & 0.30 & \textbf{0.00} & \checkmark \\
Spider-Man & 1.00 & 0.50 & \textbf{0.10} & Partial \\
Superman & 1.00 & 0.00 & \textbf{0.00} & \checkmark \\
Thor & 1.00 & 0.00 & \textbf{0.00} & \checkmark \\
Wonder Woman & 1.00 & 0.00 & \textbf{0.00} & \checkmark \\

\midrule
\textbf{Mean} & \textbf{1.00} & \textbf{0.15} & \textbf{0.02} & \textbf{98\%} \\
\bottomrule
\end{tabular}
\end{table}

Table~\ref{tab:bg_metrics_lpips} shows that \sys consistently outperforms LOCATEdit across all single-concept settings. For \emph{Violence and Weapons}, LPIPS decreases from 0.091 to 0.020, PSNR increases from 24.80 to 34.30~dB, and SSIM rises from 0.978 to 0.997.

The largest gain appears in the \emph{Multi-Concept} setting, where overlapping semantics and strong self-attention connectivity increase edit spillover. Here, \sys reduces LPIPS from 0.166 to 0.058, improves PSNR from 18.21 to 28.01~dB, and raises SSIM from 0.908 to 0.986. This supports the motivation in Section~\ref{sec:threat_model}: instance-consistent gating is most beneficial where satisfying P2 is hardest.

\mypara{\textbf{Comparison with \textsc{AntiMirror}}}
We also evaluate \textsc{AntiMirror}~\cite{das2025conceptreplacementtechniquesreally} on the celebrity subset, the only category supported by its face-parsing design. It achieves a mean background LPIPS of 0.021, PSNR of 27.08~dB, and SSIM of 0.987, compared with 0.020, 30.20~dB, and 0.995 for \sys. The comparable fidelity is expected because face parsing provides precise masks within a narrow concept class. The key distinction is coverage: \textsc{AntiMirror} is limited to celebrity faces and does not support the remaining four categories or multi-concept composition, whereas \sys provides comparable fidelity with open-vocabulary coverage.

\subsection{Category-Specific Validation}

Contrastive CLIP scoring captures holistic image--text similarity and may not reflect complete removal of localized concepts, particularly under spatially constrained edits; prior work notes that CLIP similarity can remain high after targeted suppression~\cite{das2025conceptreplacementtechniquesreally}. We therefore complement it with dedicated per-category detectors.

\subsubsection{Nudity}
We use the open-source NudeNet classifier~\cite{bedapudi2022nudenet}, restricting evaluation to labels corresponding to commonly prohibited content: exposed female breasts, exposed male and female genitalia, exposed buttocks, and exposed anus. Original images yield a mean nudity score of $70.51$, confirming strong explicit content prior to editing. After \sys, the mean score falls to $0$: no nudity is detected in any edited output.

\begin{table}[t]
\centering
\caption{GCD-based celebrity identity confidence before and after applying AntiMirror and the proposed \sys\ pipeline.}
\label{tab:gcd_results}

\setlength{\tabcolsep}{2pt}
\renewcommand{\arraystretch}{0.9}

\begin{tabular}{lccccc}
\toprule
\textbf{Celebrity} & Original & \multicolumn{2}{c}{AntiMirror} & \multicolumn{2}{c}{\makecell{\sys\\(Proposed)}} \\
\cmidrule(lr){2-2} \cmidrule(lr){3-4} \cmidrule(lr){5-6}

& \textbf{GCD Score}
& \makecell{\textbf{Best}\\\textbf{Celeb}}
& \makecell{\textbf{Spec.}\\\textbf{Identity}}
& \makecell{\textbf{Best}\\\textbf{Celeb}}
& \makecell{\textbf{Spec.}\\\textbf{Identity}} \\

\midrule
Brad Pitt      & 0.87 & 0.31 & 0.01 & 0.16 & \textbf{0.00} \\
Donald Trump   & 0.93 & 0.37 & 0.22 & 0.23 & \textbf{0.00} \\
Joe Biden      & 0.90 & 0.18 & 0.01 & 0.26 & \textbf{0.00} \\
Elon Musk      & 1.00 & 0.26 & 0.00 & 0.11 & \textbf{0.00} \\
\bottomrule
\end{tabular}
\end{table}

\subsubsection{Copyrighted Content}
\label{copycat}
We adopt the CopyCAT evaluation framework~\cite{he2024fantastic}, which uses a multimodal model (GPT-4o in the original work) to report both a general character-like detection signal and a stricter specific-identity score. Table~\ref{tab:identity_suppression} summarizes results over 100 copyrighted-character images (10 per character).

\sys reduces mean general detection from $1.00$ to $0.15$, and mean specific-identity detection to $0.02$---a $98\%$ reduction. Specific identity is eliminated entirely in 98 of 100 images; residual signal persists only for \emph{Captain America} and \emph{Spider-Man}, reflecting borderline visual similarity rather than systematic failure.

The residual general detections are themselves informative. Inspecting CopyCAT's output labels shows they correspond to \emph{generic} descriptors, ``superhero,'' ``cat character,'' ``mouse character'', rather than protected identities. \sys therefore performs \emph{identity neutralization}: it strips the distinctive, legally protected visual features while retaining the broad semantic category the user requested, leaving coherent but non-infringing content.

We note that CopyCAT evaluates one copyrighted concept at a time and does not support multiple distinct identities within a single image, whereas \sys detects and suppresses several simultaneously---a requirement for real deployments where violations co-occur.

\begin{figure}[!t]
\centering
\includegraphics[width=\columnwidth]{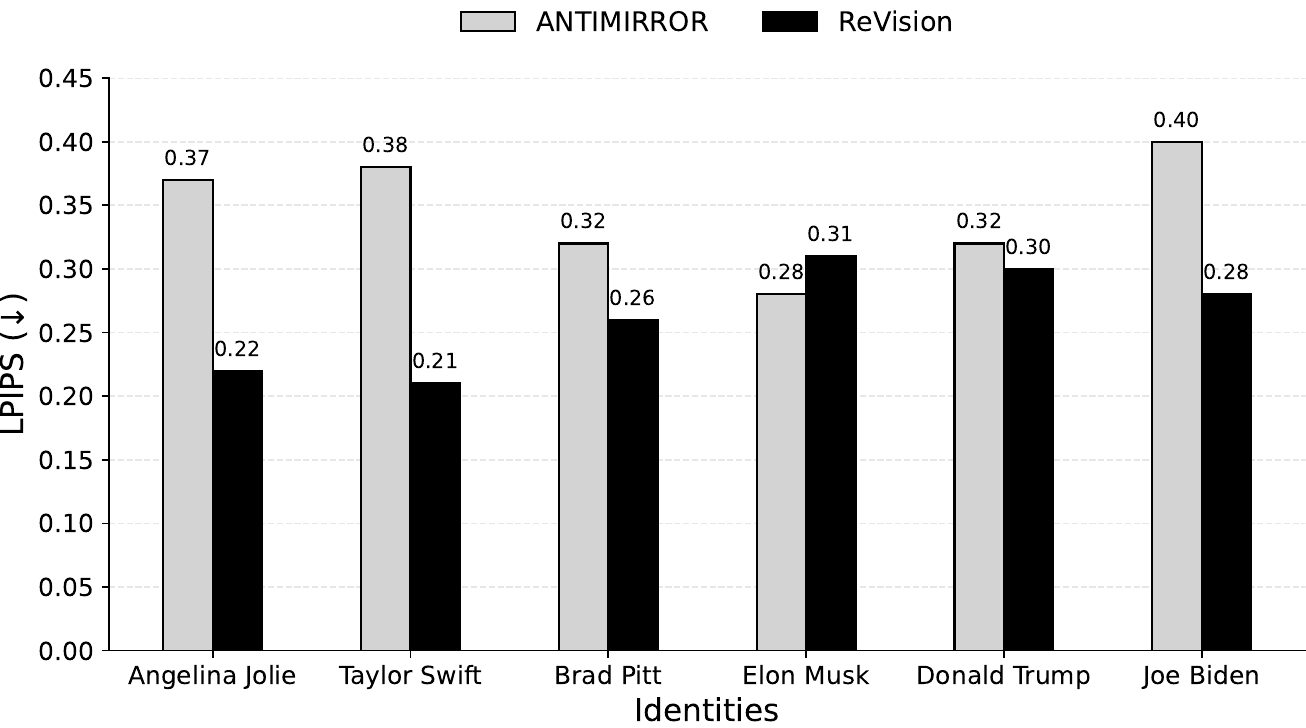}
\caption{LPIPS comparison between AntiMirror and \sys across different celebrities on the AntiMirror benchmark dataset.}
\label{fig:lpips_transposed}
\end{figure}

\subsubsection{Celebrity Identity}
To assess suppression of recognizable public figures we use the Giphy Celebrity Detector (GCD),\footnote{\url{https://github.com/Giphy/celeb-detection-oss}} which scores identity similarity against a set of known celebrities~\cite{das2025conceptreplacementtechniquesreally}. Higher scores indicate stronger confidence that a specific identity is present.
Before editing, the undefended generator reproduces highly recognizable identities ($0.87$ for Brad Pitt, $0.93$ for Donald Trump, $0.90$ for Joe Biden, $1.00$ for Elon Musk). After \sys, confidence in the targeted identity falls to $0.00$ in every case (Table~\ref{tab:gcd_results}). GCD occasionally reports non-zero confidence for \emph{alternative} identities ($0.11$--$0.26$), but these are low-confidence, ambiguous matches rather than substitution of one recognizable celebrity for another.

\begin{table}[t]
\centering
\caption{GCD comparison between AntiMirror and \sys\ on the AntiMirror celebrity benchmark dataset. Lower GCD indicates better suppression of celebrity identity. Best values are shown in bold.}
\label{tab:gcd_antimirror_clean}

\setlength{\tabcolsep}{2pt}
\renewcommand{\arraystretch}{0.9}

\begin{tabular}{lcc}
\toprule
\textbf{Celebrity} &
\textbf{AntiMirror GCD} ($\downarrow$) &
\textbf{\sys\ GCD} ($\downarrow$) \\
\midrule

Angelina Jolie & 0.23 & \textbf{0.00} \\
Taylor Swift   & 0.05 & \textbf{0.04} \\
Brad Pitt      & \textbf{0.00} & \textbf{0.00} \\
Elon Musk      & 0.03 & \textbf{0.00} \\
Donald Trump   & \textbf{0.00} & \textbf{0.00} \\
Joe Biden      & 0.05 & \textbf{0.00} \\

\bottomrule
\end{tabular}
\end{table}

\subsection{Generalization to Other Datasets}

We assess generalization on benchmarks from prior concept-removal work, examining edited-image quality and adversarial robustness; detection accuracy is reported in Section~\ref{sec:detection_accuracy}.

\mypara{\textbf{Comparison with \textsc{AntiMirror}}}
Fig.~\ref{fig:lpips_transposed} shows \sys attains lower full-image LPIPS than \textsc{AntiMirror} across all six identities---most notably $0.37 \to 0.22$ for Angelina Jolie and $0.38 \to 0.21$ for Taylor Swift---indicating better preservation of visual fidelity during identity suppression. As \textsc{AntiMirror} reports full-image LPIPS, this metric also reflects changes within edited regions. Critically, neither Jolie nor Swift appears in our benchmark, yet \sys detects and suppresses both, demonstrating generalization to unseen identities.

Table~\ref{tab:gcd_antimirror_clean} compares GCD scores on the \textsc{AntiMirror} benchmark. Here the \textsc{AntiMirror} column reports the values published by its authors on their own dataset, while the \sys column reports our results on that same dataset; this differs from Table~\ref{tab:gcd_results}, where we apply \textsc{AntiMirror} to \emph{our} benchmark. Evaluated on its home distribution, \textsc{AntiMirror} performs strongly, yet \sys matches or improves on it for every identity---reducing Angelina Jolie from $0.23$ to $0.00$, and Elon Musk and Joe Biden from $0.03$ and $0.05$ to $0.00$. Taken together, the two tables show that \sys attains comparable or better identity suppression both on \textsc{AntiMirror}'s own benchmark and on images drawn from a different generator distribution, without concept-specific machinery.

\mypara{\textbf{Adversarial robustness}}
We evaluate on MMA-Diffusion~\cite{yang2024mma}, designed to evade both prompt-based and post-hoc defenses via prompt obfuscation and image perturbation (61 source images $\times$ 4 adversarial variants $=$ 244 samples). \sys attains $81.15\%$ detection accuracy. For context, MMA-Diffusion reports an $85\%$ attack success rate against the conventional detectors it targets, implying roughly $15\%$ detection under the same attack---though we note this figure is derived from their reported ASR rather than measured by us under identical conditions. We attribute \sys's resilience to joint reasoning over image content and policy descriptions, which does not depend on the surface form of the prompt.

\begin{table}[t]
\centering
\caption{Human moderation results showing recognizability of sensitive content before and after applying \sys.}
\small
\renewcommand{\arraystretch}{1.15}
\setlength{\tabcolsep}{3pt}

\begin{tabular}{p{2.1cm}cc}
\toprule
\makecell{\textbf{Condition}} &
\makecell{\textbf{Recognizable}\\\textbf{Content (\%)}} &
\makecell{\textbf{Suppression}\\\textbf{Rate (\%)}} \\
\midrule

Original & 95.99 & -- \\

\makecell[l]{\sys\\(With Generic Label)}
 & 15.40 & 84.60 \\

\makecell[l]{\sys\\(Final)} & 10.16 & 89.84 \\
\bottomrule
\end{tabular}

\label{tab:human_eval}
\end{table}

\subsection{Human Moderation Evaluation}
\label{sec:human_eval}

To complement automated metrics, we conducted a human moderation audit to assess whether unacceptable content remains recognizable. Following prior work~\cite{he2024fantastic}, human judgment validates recognizability rather than replacing automated evaluation at scale.

We sampled a stratified subset of 36 images from the 800-image benchmark: 26 single-concept and 10 multi-concept, balanced across conditions (18 original and 18 \sys-edited). Paired pre-/post-edit versions of the same image were excluded to avoid anchoring. Using a custom web interface in a controlled lab setting, we collected annotations from 35 volunteer second-year undergraduate computer science students who were not otherwise involved in the research. All participants received informed consent and content-warning forms; participation was voluntary, and no identifying information was collected.

For each image, annotators indicated whether it contained one of five moderation-relevant categories aligned with the detector scope---(A) copyrighted fictional characters, (B) drugs or alcohol, (C) weapons or violence, (D) public figures, and (E) nudity or sexualized content---using \textit{Yes}, \textit{No}, or \textit{Unsure}. For \textit{Yes}, they also provided a 1--3 word description of the recognized entity. Images were randomized and annotated independently. An image was counted as recognized when the annotator selected \textit{Yes} for the category matching its generated concept; \textit{Unsure} was conservatively treated as a non-detection. This produced $18 \times 35 = 630$ judgments per condition and 1260 overall.
As shown in Table~\ref{tab:human_eval}, original images are recognizable in $95.56\%$ of judgments ($602/630$), confirming that sensitive content is clearly identifiable before editing. The remaining responses were primarily \textit{Unsure} cases in which annotators were unfamiliar with the depicted concept.

After \sys, recognizability falls to $15.40\%$, corresponding to an $84.60\%$ suppression rate ($533/630$). Many residual \textit{Yes} responses used generic descriptions such as `TV character'' or `superhero'' rather than naming a specific character or public figure, consistent with the CopyCAT results in Section~\ref{copycat}. Excluding these generic identifications reduces recognizability to $10.16\%$, yielding a final suppression rate of $89.84\%$ ($566/630$): in roughly nine out of ten cases, human evaluators cannot confidently recognize the protected concept in \sys-edited images.

\section{Limitations and Future Work}
\label{sec:limitations}
Our benchmark is annotated by construction: each image inherits the concept labels of its generating prompt, making these labels \emph{intended} rather than \emph{verified}.  We assume that the prompted unsafe concept appears correctly in the generated image. Since text-to-image models sometimes do not faithfully render every prompted element, a failure mode termed \emph{catastrophic neglect}~\cite{chefer2023attend} that worsens under compositional prompts~\cite{huang2023t2i}, a reported miss conflates detector failure with the generator never having rendered the concept, most notably for small held objects such as cigarettes. This effect is symmetric across all evaluated methods, so comparisons remain valid even where absolute per-category values are optimistic. A related gap concerns the bounding boxes on which gating depends: lacking human-annotated reference regions, we report no localization accuracy; our evidence that localization is adequate is therefore indirect, resting on background fidelity, qualitative comparison, and human moderation. 

\mypara{Future Work} Closing this gap requires human verification of which prompted concepts are actually \emph{visible}, followed by instance-level reference boxes enabling IoU-based scoring, which would in turn permit a systematic comparison of open- and closed-source VLMs as grounded safety detectors separating three abilities that current evaluations conflate: whether a model recognizes a concept, whether it localizes it correctly, and whether it will answer at all. Beyond evaluation, further directions include multi-edit diffusion methods that edit several concepts simultaneously to reduce latency in complex scenes~\cite{zhu2025mde}, and robustness against adversarial attacks on VLM-based detectors~\cite{guo2024efficient}, including attack transferability from open- to closed-source VLMs under realistic deployment~\cite{schaeffer2024failures}.

\section{Conclusion}

We presented \sys, a training-free, prompt-driven, post-hoc safety framework for visual generative models that detects policy-violating concepts and performs targeted replacement during inference. By combining VLM-based semantic auditing with instance-consistent spatial gating, \sys addresses a key usability failure of post-hoc editing: imprecise spatial localization in complex scenes. Through extensive evaluation, we show that \sys reliably suppresses unsafe concepts while preserving non-target visual regions, as measured by background fidelity, semantic alignment, category-specific detectors, and human moderation. Unlike approaches that require retraining, rely on concept-specific masks, or focus on isolated violations, \sys operates modularly as a provider-side last-line defense and scales across diverse safety categories via policy prompts under realistic black-box auditing assumptions. Overall, our results show that practical safety enforcement in generative systems requires not only detecting unsafe content, but also preserving benign content to maintain output integrity.

\section*{Acknowledgements}
This project was supported by the Center of Excellence in Data Science and Artificial Intelligence, a collaboration between Thapar Institute of Engineering and Technology (TIET), Patiala, and The University of Queensland.
We also thank the Thapar School of Advanced AI and Data Science, TIET, Patiala, for providing access to their computational resources.

\bibliographystyle{ACM-Reference-Format}
\bibliography{references}
\appendix

\section{Detector Ablation}
\label{app:detector_ablation}

\sys's Stage~1 interface is prompt-driven and not tied to a specific VLM. To characterize how detection quality varies with the underlying model, we instantiate the identical policy prompt and output schema with two alternatives---GPT-4o and the open-weights Qwen2.5-VL-7B-Instruct---and evaluate all three on the full 800-image benchmark, with no difference in prompt, parsing, or scoring code.

Table~\ref{tab:detector_ablation} reports per-image category-level detection accuracy under the protocol of Section~\ref{sec:detection_accuracy}. Overall accuracy is comparable ($88.4\%$, $84.5\%$, $83.1\%$), but the per-category profile differs sharply and no detector dominates: Gemini-3.5-Flash leads on public figures and weapons, GPT-4o on smoking and alcohol, and Qwen2.5-VL-7B reports by far the highest nudity figure.

That nudity result illustrates why these figures need care. The metric is recall against prompt-intended labels, so it rewards over-flagging and cannot separate a correct detection from a false positive. Manual inspection found that many male-nudity images depict a clothed or only partially unclothed subject---the prompted concept was not faithfully rendered. The closed-source detectors' lower scores may therefore reflect correct restraint rather than failure. Lacking human-annotated labels (Section~\ref{sec:limitations}), we cannot adjudicate this, so we report the figures uncorrected and draw no ranking from this category.

The public-figure result is clearer. GPT-4o declines to respond on $41$ of $800$ images, $40$ containing public figures; counted as non-detections, these reduce its accuracy from $78.2\%$ over attempted images to $61.5\%$ overall. Qwen2.5-VL-7B never refuses but often fails to recognize less prominent individuals. The two reach nearly identical scores ($61.5\%$, $60.4\%$) through different mechanisms---one a policy decision, the other a capability limit---and only the former is visible in the response and recoverable by the provider.

We report this to characterize a dependency, not to rank detectors. Detection quality is a design axis separate from our contribution: gating consumes a bounding box irrespective of its source, and Sections~\ref{fidelity-sec}--\ref{sec:human_eval} establish its effectiveness on the instances that are detected. The ablation bears on which detector a deployment should select---possibly a category-dependent choice---not on whether gating works once a concept is located.

\begin{table}[t]
\centering
\caption{Per-image category-level detection accuracy (\%) on the 800-image benchmark, with the Stage~1 interface instantiated using three detectors. The metric is recall against prompt-intended labels and does not penalize false positives; the nudity column should not be read as a ranking (see text and Section~\ref{sec:limitations}). Refusals count as non-detections.}
\label{tab:detector_ablation}
\small
\setlength{\tabcolsep}{4pt}
\renewcommand{\arraystretch}{1.05}
\begin{tabular}{@{}lccc@{}}
\toprule
\textbf{Category} &
\makecell{\textbf{Gemini}\\\textbf{3.5-Flash}} &
\textbf{GPT-4o} &
\makecell{\textbf{Qwen2.5}\\\textbf{VL-7B}} \\
\midrule
Copyrighted content & 99.8 & \textbf{100.0} & 96.4 \\
Public figures      & \textbf{82.9} & 61.5 & 60.4 \\
Nudity              & 59.3 & 74.1 & 98.2 \\
Smoking \& alcohol  & 60.4 & \textbf{69.3} & 67.3 \\
Violence \& weapons & \textbf{93.4} & 83.9 & 79.6 \\
\midrule
\textbf{Overall}    & \textbf{88.4} & 84.5 & 83.1 \\
\bottomrule
\end{tabular}
\end{table}


\section{Deployment Considerations}
\label{app:deployment}
Post-hoc safety mechanisms add inference-time latency, since they operate after image generation. In our setup Stable Diffusion~3.5 Turbo and FLUX.1-dev define the baseline generation latency, and \sys adds overhead from Stage~1 semantic analysis and Stage~2 localized editing.

Stage~1 latency depends on the detector. Gemini-3.5-Flash requires a median of $7.2$\,s per image and GPT-4o $8.2$\,s, both dominated by API round-trip; the GPT-4o figure was measured under concurrent load subject to provider rate limiting and is therefore an upper bound. The open-weights Qwen2.5-VL-7B, served locally with vLLM on a single H100, is markedly faster at $1.6$\,s median, incurring no network round-trip and admitting local batching. This exposes a deployment trade-off: hosted detectors supply the grounding quality on which \sys's results depend (Appendix~\ref{app:detector_ablation}), but impose per-image latency that dominates the pipeline, whereas a co-located open model is roughly $5\times$ faster at a substantial accuracy cost. For providers already operating GPU infrastructure for generation, co-locating the detector is the more practical configuration if detection quality can be improved.

Stage~2 adds $1$--$2$\,s per concept, applied sequentially, so a four-concept image incurs $6$--$8$\,s of editing. \sys uses a lightweight diffusion model (Stable Diffusion~1.5) for editing rather than the base generator, reducing memory and compute cost, though this may occasionally introduce minor stylistic inconsistency. Multi-edit diffusion methods that revise several concepts in a single pass~\cite{zhu2025mde} would remove this sequential dependence.

\end{document}